\documentclass[conference]{IEEEtran}

\usepackage{cite}
\usepackage{amsmath,amssymb,amsfonts}
\usepackage{algorithm}
\usepackage{algorithmic}
\usepackage{graphicx}
\usepackage{textcomp}
\usepackage{url}
\usepackage{xcolor}
\usepackage{booktabs}
\usepackage{cleveref}
\usepackage{enumitem}
\usepackage{pifont}
\usepackage{listings}
\usepackage{multirow}
\usepackage{tikz}
\usepackage{tcolorbox}

\NewDocumentEnvironment{alignb}{b}{%
  \begin{align*}
  \refstepcounter{equation} #1 \tag{\theequation}
  \end{align*}
}{\ignorespacesafterend}

\def\BibTeX{{\rm B\kern-.05em{\sc i\kern-.025em b}\kern-.08em
    T\kern-.1667em\lower.7ex\hbox{E}\kern-.125emX}}
\begin{document}

\title{Integrating Performance Tools in Model Reasoning for GPU Kernel Optimization}

\author{\IEEEauthorblockN{Daniel Nichols\IEEEauthorrefmark{2}, Konstantinos Parasyris\IEEEauthorrefmark{2}, Charles Jekel\IEEEauthorrefmark{2}, Abhinav Bhatele\IEEEauthorrefmark{4}, Harshitha Menon\IEEEauthorrefmark{2}}
\IEEEauthorblockA{~\\
    \textit{\IEEEauthorrefmark{2}Lawrence Livermore National Laboratory}\\
    \textit{\IEEEauthorrefmark{4}University of Maryland}\\
    E-mail: \IEEEauthorrefmark{2}danielnichols@llnl.gov, \IEEEauthorrefmark{2}harshitha@llnl.gov}
}

\maketitle

\begin{abstract}
Language models are now prevalent in software engineering with many developers using them to automate tasks and accelerate their development. While language models have been tremendous at accomplishing complex software engineering tasks, there are still many areas where they fail to deliver desirable results, for instance code performance related tasks. Tasks like optimization depend on many complex data from the environment, hardware, etc. that are not directly represented in source code. Recent efforts have seen large improvements in general code modeling tasks using chain-of-thought style reasoning, but these models still fail to comprehend how the environment interacts with code performance. In this paper we propose a methodology to train language models that can {\it interact} with performance tools during their reasoning process. We then demonstrate how this methodology can be used to train a state-of-the-art GPU kernel optimization model.
\end{abstract}


\section{Introduction}

Language models (LMs) are now routine parts of software development workflows. They accelerate everyday programming tasks and increasingly tackle nontrivial code generation and refactoring problems~\cite{chen2021evaluating,li2022competition,roziere2023code,wang2021codet5,wang2023codet5plus,zheng2023codegeex}. Despite this progress, performance optimization remains a weak spot: optimizing code requires knowledge of details that lie outside source code, such as runtime behavior, compiler decisions, and hardware effects that only show up when code runs on a specific system. Current text-only models cannot properly evaluate how the code they generate will interact with computer hardware~\cite{pyda_shortcomings_2025}. Even agentic systems that can interact with the programming environment still struggle to solve even basic correctness bugs at the repository level~\cite{jimenez_swe-bench_2023,rashid_swe-polybench_2025}.

A particularly challenging optimization problem that is becoming increasingly relevant is GPU kernel optimization. There are many competing factors that affect kernel performance such as  register pressure, occupancy, memory access patterns, compiler scheduling, etc. While some of these kernel properties can be loosely inferred from the source code, it remains a complex engineering problem to reason about how kernels will perform on GPU hardware. Even human developers have to consult external tools, such as compiler reports, profilers, and targeted microbenchmarks, to form and test hypotheses. Using LMs to optimize kernels without similar access to these optimization toolchains is futile; without dynamic environment feedback the model is left to guess and use learned heuristics to optimize code performance. Many of these tools are complex and difficult to effectively utilize whether by human engineers or state-of-the-art LMs. Furthermore, once a model has access to these data sources it must be able to effectively use their outputs. For example, knowing a kernel has 14 register spills does not assist the model unless it has been properly trained to utilize this information. 

We choose to focus on GPU kernels as they are becoming increasingly relevant in the high performance computing and machine learning landscapes and provide a well-defined, scoped optimization problem to study. GPU kernels are often the compute intensive portions of scientific and machine learning applications, thus small kernel-level gains translate into meaningful end-to-end savings at scale. This makes kernels an ideal application target, as optimizations can have a meaningful effect on time to results. For example, in a case study presented later in this work, our model optimizes an expensive kernel in a shock-hydrodynamics application, leading to a measurable reduction in total runtime on an NVIDIA H100 system.

Training reasoning and tool calling models is, however, non-trivial. Recent works have found verifiable rewards to be vital in employing reinforcement learning (RL) to fine-tune models to particular output distributions~\cite{deepseek-ai_deepseek-r1_2025}. This requires training data that has verifiable rewards; in the context of code performance this is difficult due to the scarcity of easily runnable data and the infrastructure required to run kernel benchmarking during training. Furthermore, RL fine-tuning requires significant compute resources. Recent techniques, such as GRPO, require running the model in inference (e.g. generating samples) during training, which dramatically slows down training~\cite{hpc_r1:sc25}. In addition, works have found that sufficiently large models ($>34$B parameters) are required to get GRPO to converge meaning tens to hundreds of GPUs will be needed for fine-tuning~\cite{deepseek-ai_deepseek-r1_2025}. Fine-tuning models to call performance tools presents another bottleneck on top of these standard computational burdens. The training process frequently has to wait on long tool calls before it continues.
Due to the immense data and compute bottlenecks it is difficult to design and iterate on the creation of a reasoning LM for code optimization.

We address these challenges by training a reasoning LM that interacts with performance tools during inference. Concretely, the model receives an optimization task to optimize a GPU kernel in a larger application and iterates on hypotheses, benchmarking, and profiling in a long reasoning chain until it finds a suitable, correct optimization. We fine-tune the model with a reinforcement-learning setup designed to encourage tool use and verifiable optimization performance, and we demonstrate a compact toolset that enables models to effectively interact with the runtime environment to optimize code. 

In this paper we make the following important contributions:

\begin{itemize}
    \item We introduce a novel approach to fine-tune language models to call performance tools, reason about performance, and optimize code. Using this approach we fine-tune the opt-r1 and opt-r1-mini models. %
    \item We evaluate our approach using state-of-the-art language models and demonstrate strong optimization performance across benchmark kernels.
    \item We demonstrate how to distill optimization reasoning into small language models that are easy to run, but achieve similar performance to larger optimization models.
    \item We optimize a GPU kernel in a real HPC application that yields a 17\% kernel speedup and 3\% end-to-end speedup for the total application. %
\end{itemize}

\section{Background}
In this section we present background on how reinforcement learning (RL) can be deployed to fine-tune an existing language model to a particular output distribution.
We further detail how test-time-compute, i.e. letting the model spend more compute during inference, can be utilized to drastically increase the ability of LMs to do complex problem solving.

\subsection{Fine-tuning LMs with RL}

After a LM has been trained on a large corpus of text data it is usually desirable to further {\it align} it with a particular output distribution.
For example, a LM just trained on a big corpus of text will learn the distribution of text in its dataset well, but will have poor performance on out of distribution tasks such as multi-turn dialogue with a user.
If the training text originates from articles, books, websites, etc., then the LM will output text according to the distribution of those documents.
For this reason, after LMs are trained on a large pre-training dataset they are then fine-tuned on a smaller dataset of instruction-response and/or dialog data to enable them to engage in dialogue with a user.
They retain much of the learned language modeling capabilities from their pre-training, but will now output in an instruction following, dialogue oriented manner.
This has become known as instruction tuning~\cite{zhang_instruction_2025}.

LMs can be fine-tuned to behaviors more complex than just instruction following. Many modern models are fine-tuned with variations of RL to have desirable output properties such as safety, morality, etc.
For example, reinforcement learning with human feedback (RLHF) is a popular RL fine-tuning technique to align a model's outputs with those labeled as more desirable by human annotators~\cite{ouyang_training_2022}.
Many of these alignment techniques focus on using RL to optimize the model weights.
There are now many techniques in the literature for using RL to fine-tune LMs; RLHF and PPO were the original popular approaches and now Group Relative Policy Optimization (GRPO) is behind most new LM RL fine-tunings~\cite{shao_deepseekmath_2024,deepseek-ai_deepseek-r1_2025}.

A general setup for RL on LMs defines a policy $\pi_\theta(x \mid c)$ over model outputs $x$ given context $c$ (prompt, system instructions) and a reward function $r(x)$ that signals the behavior we want the model to learn. During training, the model generates completions to some training problems, receives rewards based on its outputs, and updates its weights $\theta$ to increase the likelihood of generating high reward outputs again. Typically, a copy of the original model, called the reference model, is used for regularization to prevent the model being fine-tuned from moving too far away from its original distribution. 
GRPO follows this same RL approach, but uses grouped relative scores, or ``advantages", instead of direct rewards. In this setting, the LM generates $K$ completions to an input sample problem where $K$ is the group size. The reward for each completion is computed individually, but their final reward is a normalized score relative to the rest of the completions in the group. This gives the model information about good and bad outputs and how much of an advantage certain outputs have over others.
The learning objective for GRPO is shown in \Cref{eq:bg:grpo}.

\begin{alignb}\label{eq:bg:grpo}
    \mathcal{L}_{\text{GRPO}}
    = & \mathbb{E}_{x \sim \pi_\theta(\cdot \mid c)}
    \Big[ -w(x)\log \pi_\theta(x \mid c) \Big] \\
    & + \beta \mathrm{KL}\left(\pi_\theta(\cdot \mid c) | \pi_{\text{ref}}(\cdot \mid c)\right)
\end{alignb}

Here $w(x)$ is a group-normalized value of $r(x)$ (i.e. the ``advantage" of $x$ over other generations in the group), and $\beta$ controls regularization.

\noindent\textbf{Verifiable Rewards}
The reward function in RL can be any function that yields higher outputs for the type of data you want to incentivize.
In initial works on RL fine-tuning of LMs it was typical to train a separate LM as the reward model; this reward LM would usually score outputs based on human preference.
This produces a very noisy reward signal, encouraging {\it reward hacking}, where the model being fine-tuned learns how to ``cheat" the reward LM to get higher rewards.
For this reason, verifiable reward functions have become common in RL fine-tuning.
A verifiable reward is one that can be directly computed, such as length of the generated output, rather than one that has to be predicted by a model.
A common training dataset used for verifiable reward fine-tuning is GSM8K where the model is tasked with solving a math problem.
The reward function is easy to compute here as we simply need to verify if the model outputs the correct number or not~\cite{cobbe_training_2021}.

\subsection{Test-time compute and LM Reasoning}

Test-time compute (TTC) refers to spending additional computation at inference to improve answer quality~\cite{snell2024scaling,madaan2023self,brown2024large}. Common TTC techniques include sampling multiple candidates and selecting with a reward mode (best-of-n sampling), tree-style exploration over intermediate thoughts, and chain-of-thought~\cite{cot} to allow the model to internally iterate and reason on its answers. TTC is a trade-off between compute and high quality outputs; generally spending more inference time will yield better outputs from the model but with longer turnaround times and higher costs. 

\paragraph{Chain-of-thought (CoT) and Reasoning}
Allowing a model to generate intermediate reasoning (“thoughts”) can considerably improve how it handles complex reasoning and/or multi-stage tasks~\cite{cot}. Consider a simple but relevant example: choosing a GPU kernel launch configuration. A non-reasoning model might output a guessed block size; if similar prompts were in its training data, then the model might guess the best block size, but never reconsider or evaluate its own guess. A CoT-style model will instead deconstruct its answer into step-by-step reasoning: for example, (1) estimate arithmetic intensity and memory pressure from the kernel body; (2) hypothesize that occupancy is limited by registers; (3) propose a new block size; (4) analyze how the kernel will interact with the new block size; (5) iterate if needed; (6) output and explain its final answer. This process is even more successful if the model is able to call outside tools to verify hypotheses, as shown later in this work. 

CoT reasoning is typically achieved by fine-tuning models with RL to output their thoughts as they solve problems. During the fine-tuning process the models are tasked with solving problems (e.g. math puzzles) and presenting their reasoning as they arrive at a solution. They are then rewarded based on how well they can solve the problems. Even without directly rewarding the presence of reasoning, this fine-tuning process tends to yield long reasoning behavior in model and drastically improves their capabilities on more complex tasks. The reasoning simply becomes intermediate text that the model generates first before it generates its final answer. This intermediate text is often not perfectly human-like, however, the explicit stepwise structure allows the model to revisit and critique previous hypotheses, while also deconstructing problem solving into smaller, simpler components.

\paragraph{Cost–quality trade-offs}
TTC allows users to spend more compute in order to get better results.
However, it has diminishing returns: more reasoning steps or deeper searches will eventually saturate. The model will either reach its context token limit or the limits of its fundamental reasoning capabilities. For example, even with no time limits a model will reach its context window limit before arriving at a solution if tasked with solving an $\mathcal{NP}\text{-hard}$ problem with a large input size. Despite this limit we can still achieve significant improvements on complex tasks by enabling models to reason through problems before outputting answers.

\section{Overview}
In this section we present an overview of the data collection, methodology, training, and evaluation presented in this paper in order to create and study a reasoning LM that dynamically interacts with code performance during its reasoning.

Training a reasoning capable LM with GRPO requires a large dataset with verifiable rewards.
For this work that means thousands of GPU kernel optimization problems with benchmarking code and correctness validation.
To collect this data we employ a mix of synthetic data generation and distilling kernels from the HeCBench benchmark (see \Cref{sec:data-collection}).
We use this dataset to fine-tune a 70b parameter LM to optimize GPU kernels and reason about their optimization with tool calls (see \Cref{sec:model-design,sec:fine-tuning}).
This allows models to first reason about code performance and test hypotheses on the actual hardware before proposing code edits.
Finally, we use output traces from the 70b model to fine-tune a 8b parameter LM for reasoning and GPU kernel optimization.

Evaluations on realistic GPU kernels distilled from scientific applications and common algorithms are used to evaluate the two optimization models (see \Cref{sec:results}).
We analyze how well they can optimize kernels, what kinds of optimizations they are making, and how effective tool calling is in the reasoning process.
Finally, we demonstrate the effectiveness of the models on a real application kernel and show an improvement in total application runtime (see \Cref{sec:case-study}).

\section{GPU Kernel Data Collection}\label{sec:data-collection}
Large scale GRPO fine-tuning has significant data requirements in terms of volume and structure. In this section we detail the fine-tuning dataset design and requirements. We further highlight how we collect realistic benchmark kernels at scale for use during fine-tuning.

\subsection{Dataset Requirements}\label{sec:data:requirements}
Unlike standard LM supervised fine-tuning or even older RL techniques like RLHF, GRPO has unique requirements in terms of data {\bf volume} and {\bf structure}.
Using RL for fine-tuning LMs often requires thousands to tens of thousands of training iterations.
This, in turn, means that we need thousands to tens of thousands unique data samples for fine-tuning.
An effective dataset for fine-tuning a LM for GPU kernel optimization requires on the order of ten-thousand kernels.
Furthermore, recent literature has found that verifiable rewards lead to significantly better training results when doing RL on LMs~\cite{shao_deepseekmath_2024, deepseek-ai_deepseek-r1_2025}.
This means our tens of thousands of GPU kernel samples need to have {\it correctness} and {\it performance} tests so that we can verify if LM generations do in fact produce valid performance optimizations.

Collecting tens of thousands of GPU kernels with infrastructure for testing and benchmarking them is non-trivial.
The few viable candidate kernel sets we found are all in the order of hundreds of kernels and were ideal {\it benchmarks} to evaluate our fine-tuned model; it is imperative to avoid fine-tuning on the kernels we use for evaluation to avoid overfitting.
To overcome this difficulty we ameliorate a large open-source GPU kernel benchmark, HeCBench, with several transformations to increase data samples (see \Cref{sec:data:hecbench}).
We further expand the dataset by adapting synthetic data creation techniques from related recent works~\cite{nichols_performance-aligned_2024, chaturvedi_hpc-coder-v2_2024, wei_magicoder_2024} to generate more viable kernel optimization problems (see \Cref{sec:data:synthetic}).

\subsection{HeCBench GPU Kernels}\label{sec:data:hecbench}

To construct our dataset we first employ the HeCBench~\cite{jin_benchmark_2023} benchmark to obtain a broad set of runnable GPU kernels.
HeCBench was originally designed to study the performance and portability of SYCL kernels, but due to its vast library of application-inspired CUDA, HIP, OpenMP, and SYCL kernels it has been a very useful data source for many HPC studies.
There are 228 total kernels in the benchmark that vary from core algorithms, such as matrix multiplication, to application driven kernels, such as compression and multigrid.

We are able to expand the number of optimization problems by considering different kernels and/or sets of kernels from each problem.
Each benchmark kernel in HeCBench consists of a couple files: source code, makefile, etc.
The source files typically contain several CUDA kernels and device functions.
In our problem setup (later described in \Cref{sec:model-design}) we task the LM with optimizing a region of source code, which may contain several kernels or functions.
This allows us to transform a kernel benchmark in HeCBench that has two kernels, A and B, into three optimization problems: optimize A, B, and A+B together.

\subsection{Synthetic Kernels}\label{sec:data:synthetic}

We further expand on the dataset of HeCBench kernels with a dataset of synthetically generated GPU kernels and drivers.
While synthetically generated code samples pose the danger of introducing noise into the training process, many recent works have seen tremendous success from using them in training~\cite{nichols_performance-aligned_2024, chaturvedi_hpc-coder-v2_2024, wei_magicoder_2024}.
Following this trend, we use gpt-4o to generate a 5k sample data set of GPU kernels.
We provide an existing driver and Makefile structure to the LM and a random snippet of code from The Stack dataset~\cite{kocetkov_stack_2022} as inspiration to promote diversity in its outputs; the LM is utilized solely to write the kernel and kernel launch portions of the code.
This is particularly well-suited to our use case because we only need the LM to generate a runnable kernel; it does not need to be fast and is actually a more helpful data sample if it is not as it provides a better optimization problem.
After generating the 5k synthetic samples we then filter by those that actually compile and run; kernels that did not compile or run were mostly due to the model misunderstanding the template and trying to generate driver functions that already existed.
This filtering step left $\approx$ 4.9k samples for our final dataset.

\section{Model Design: Reasoning with Optimization Tools}\label{sec:model-design}
\definecolor{reasoningcolor}{RGB}{100, 150, 200}
\definecolor{toolcolor}{RGB}{150, 100, 200}
\definecolor{resultcolor}{RGB}{100, 200, 150}
\definecolor{finalcolor}{RGB}{220, 100, 100}

\newcommand{\reasoning}[1]{\tikz[baseline=(text.base)]{\node[fill=reasoningcolor!20, rounded corners=2pt, inner sep=2pt, text=reasoningcolor!80!black] (text) {\texttt{\small #1}};}}
\newcommand{\tool}[1]{\tikz[baseline=(text.base)]{\node[fill=toolcolor!20, rounded corners=2pt, inner sep=2pt, text=toolcolor!80!black] (text) {\texttt{\small #1}};}}
\newcommand{\result}[1]{\tikz[baseline=(text.base)]{\node[fill=resultcolor!20, rounded corners=2pt, inner sep=2pt, text=resultcolor!80!black] (text) {\texttt{\small #1}};}}
\newcommand{\final}[1]{\tikz[baseline=(text.base)]{\node[fill=finalcolor!20, rounded corners=2pt, inner sep=2pt, text=finalcolor!80!black] (text) {\texttt{\small #1}};}}

\newcommand{\withrowspacing}[2]{\begingroup\renewcommand{\arraystretch}{#1}#2\endgroup}

In this section we describe how the reasoning model we fine-tune interacts with code, tools, and a build/benchmark harness to iteratively optimize GPU kernels. 
The model is a standard reasoning LM assistant that receives text instructions, generates reasoning text as it tries to arrive at a solution, and finally outputs a final response as text. 
In this particular case the model is designed to receive a GPU optimization problem as input, reason about its performance and call performance tools as necessary, and finally output an optimized kernel.
The optimization target is a fixed region of interest (ROI) in a repository; the model may read the entire codebase but may only modify lines in the ROI.

\subsection{Model Inputs, Outputs, and Reasoning}

\noindent\textbf{Inputs} At the start of each run, the model receives: (i) a system instruction that defines optimization goals and constraints (execution platform, correctness requirements, compilation target), (ii) the program path and ROI span (file and line range), (iii) the exact ROI text to be improved, (iv) a baseline execution time for the unmodified program, and (v) a catalog of available tools with their arguments and return formats. Although only the ROI may be edited, all tools operate over the full repository to allow the model to build, run, and inspect the application in context. This formulation allows the model to focus on a single feasible optimization problem, while receiving the necessary performance context from the rest of the repository.

\noindent\textbf{Reasoning} The model outputs its reasoning as it attempts to optimize the ROI. 
During reasoning it can issue tool calls (e.g., compile/benchmark, inspect compiler reports, search the codebase) and immediately use returned results (timings, register counts, assembly hints, search hits) to inform its reasoning. 
The reasoning continues until the model identifies an optimization that measurably improves the baseline while preserving correctness.
It is ultimately up to the model to decide when to stop reasoning and output its final solution.
While chain-of-thought and reasoning models have shown immense improvement over standard LMs in complex tasks, a major interpretability limitation is that their reasoning traces need not be fully human interpretable.
The models are trained based on how well they can solve problems, but not necessary provide human-like reasoning.
However, in our experiments we found the reasoning traces to be not too far from human-like optimization passes.

\noindent\textbf{Outputs} When the model has found a sufficient optimization, it emits a structured result (as {\it JSON} text) containing: (i) \texttt{final\_code}, a self-contained replacement for the ROI (no changes outside the specified region), and (ii) \texttt{explanation}, a text statement summarizing the input GPU kernel, the proposed optimization, and why the optimization improves performance. 

\subsection{Tools Available to the Model}
We expose a small, purpose-built toolset that covers typical actions required for GPU kernel optimization. 
The tools aim to provide context to the model about the source code repository and dynamic performance information about how the code interacts with and runs on the current GPU architecture.
The tool schemas (descriptions of tool, input arguments, and output) are provided to the model in its system prompt; the model can call tools by emitting a JSON schema and a special tool call token.
The inference software generating outputs from the LM has to handle parsing and dispatching tool calls from the LM once it is up and running.
Below we summarize the available tools and their typical use:

\begin{itemize}
\item \textbf{\texttt{benchmark\_code}}: Patch the target region with a proposed replacement, build the full application with optional \texttt{compile\_args}, execute the user’s run script, and return aggregate timing(s) and a correctness flag. This is the primary mechanism for validating any change; the model is expected to call it after each substantive edit.
\item \textbf{\texttt{microbench\_code}}: Create a standalone test in a single file to quickly evaluate a hypothesis in isolation (block size, memory layout, prefetch distance) without building and running the full application.
\item \textbf{\texttt{compiler\_analysis}}: Compile the current ROI and return low-level diagnostics such as register count, spill bytes, estimated occupancy drivers, and PTX/SASS excerpts. Useful for verifying whether an edit reduces register pressure or improves occupancy limits.
\item \textbf{\texttt{file\_viewer}}: Read chunks of files to inspect helper functions, type definitions, or call sites that influence kernel behavior (e.g., launch parameters, strides).
\item \textbf{\texttt{list\_dir}}: Similar to \verb|ls| shell utility. Allows the model to see files in the repository; mainly helpful as a pair to the \texttt{file\_viewer} tool.
\item \textbf{\texttt{search}}: Find substrings or regex matches across the repository (e.g., to locate kernel launches, macros controlling tile sizes, or global compile flags).
\end{itemize}

\noindent
\emph{Scope and constraints} While tools operate over the entire repository, edits are restricted to the ROI lines. This is so the model can focus on the core optimization problem while still getting context from the surrounding code. The tools do enable the model to set compiler flags as they are an important part of the code optimization process. For safety reasons, the tools do not allow the model to execute shell commands or write arbitrary code outside of the observed kernel. To ensure complete safety, the tools either need to be executed in a containerized environment or require user review and approval on each piece of LM generated code before execution.

\subsection{An Example Trace}

Trace~\ref{tab:example-trace} shows an abbreviated, stylized trace illustrating the reasoning traces the model outputs. It demonstrates how the model can form hypotheses, test them, discard unhelpful ideas, and iterate until it produces a valid improvement. For brevity, we omit full tool arguments and show summarized text.

\begingroup
\setlength{\tabcolsep}{4pt}
\renewcommand{\arraystretch}{2.25}
\begin{small}
\begin{table}[ht]
\renewcommand{\tablename}{TRACE}
\caption{Example reasoning trace.\label{tab:example-trace}}

\begin{tabular}{@{}ll@{}}
\reasoning{<|reasoning|>} & \withrowspacing{1.0}{\color{reasoningcolor} \begin{tabular}[c]{@{}l@{}}Suspect uncoalesced AoS loads; \\ try vectorize + \texttt{\_\_restrict\_\_}.\end{tabular}} \\

\tool{<|tool|>} & {\color{toolcolor}\texttt{file\_viewer(file="..", lines="..")}} \\ 

\reasoning{<|reasoning|>} & {\color{reasoningcolor} AoS confirmed; proceed with vectorized load variant.} \\ 

\tool{<|tool|>} & \withrowspacing{1.0}{\color{toolcolor}\begin{tabular}[c]{@{}l@{}} \texttt{microbench\_code(} \\ \texttt{  snippet="float4 lds + \_\_restrict\_\_",} \\ \texttt{  launch=\{block:256\})} \end{tabular}} \\ 

\result{<|result|>} & {\color{resultcolor}\texttt{time\_ms=3.91}} \\ 

\tool{<|tool|>} & \withrowspacing{1.0}{\color{toolcolor}\begin{tabular}[c]{@{}l@{}} \texttt{compiler\_analysis(config=} \\ \texttt{  \{arch:"sm\_90a",ptxas:"-v"\})} \end{tabular}} \\ 

\result{<|result|>} & {\color{resultcolor}\texttt{regs: 80→72, spills: 0B}} \\

\tool{<|tool|>} & \withrowspacing{1.0}{\color{toolcolor}\begin{tabular}[c]{@{}l@{}} \texttt{benchmark\_code(patch="A",} \\ \texttt{  compile\_args=["-O3","-arch=sm\_90"])} \end{tabular}} \\

\result{<|result|>} & {\color{resultcolor}\texttt{correct=true, time\_ms=4.22}} \\

\reasoning{<|reasoning|>} & \withrowspacing{1.0}{\color{reasoningcolor}\begin{tabular}[c]{@{}l@{}} Add \texttt{\_\_launch\_bounds\_\_(256,2)} \\ \,\, raise residency without spills. \end{tabular}} \\

\tool{<|tool|>} & \withrowspacing{1.0}{\color{toolcolor}\begin{tabular}[c]{@{}l@{}}\texttt{benchmark\_code(patch="A + } \\ \texttt{  \_\_launch\_bounds\_\_(256,2)")} \end{tabular}} \\

\result{<|result|>} & {\color{resultcolor}\texttt{correct=true, time\_ms=4.17}} \\

\final{<|final|>} & \withrowspacing{1.0}{\color{finalcolor}\begin{tabular}[c]{@{}l@{}} Emit \texttt{\{"final\_code": "...",} \\ \texttt{  "explanation": "..."\}}. \end{tabular}} \\
\end{tabular}
\end{table}
\end{small}
\endgroup

\addtocounter{table}{-1}

\section{Fine-tuning Optimization Reasoning Models}\label{sec:fine-tuning}
In this section we present our approach to employing RL to fine-tune a reasoning optimization model that can call tools during its reasoning process. We discuss how to encourage tool use in the model and reward higher performance kernels. We then discuss how a large trained model can be utilized to train a smaller reasoning optimization model. Finally, we discuss our computational approach for fine-tuning.

\subsection{GRPO Adaptations}

We build on Group Relative Policy Optimization (GRPO)~\cite{shao_deepseekmath_2024}, which optimizes the log-likelihood of sampled completions within a group while regularizing toward a reference model. We maintain the traditional GRPO policy while modifying the reward function to our use case. To incentivize using the tools and optimizing the kernels we employ a two part reward function as shown in \Cref{eq:grpo:reward}.

\begin{equation}\label{eq:grpo:reward}
  r(x) = r_{\text{perf}}(x) + r_{\text{tools}}(x)
\end{equation}

\paragraph{Performance reward}
Let $t_\text{base}$ be the baseline end-to-end time and $t(x)$ the time after applying the code change from model output $x$ (as reported by the benchmark tool). We define

\begin{equation}
r_{\text{perf}}(x)=
    \begin{cases}
        -2 & \text{if incorrect}, \\
        -1 & \text{if correct but slower}, \\
        \min\left\{10, \tfrac{t_\text{base}}{t(x)}-1\right\}
        & \text{otherwise,}
    \end{cases}
\end{equation}

so correct, fast improvements are rewarded and incorrect edits are penalized. This is the most important part of the reward function as our end goal is to optimize GPU kernels.

\paragraph{Tool-use reward}

To encourage the model to use the tools we further include a reward that promotes tool use.
Since it is likely that some of the tools may not be available when the model is being used or the model is being deployed with new tools, we want to prevent the model from overfitting to a fixed set of tools. To accomplish this we select a random set of 4 to 6 of the tools to be available for each fine-tuning sample. We ensure the \texttt{benchmark\_code} tool is always available. This allows the model to learn to use whatever tools are listed as available in its system prompt and not overfit to a provided set of tools in the fine-tuning dataset. Thus, we use a scaled reward based on the percentage of tools available that are used, as shown in \Cref{eq:grpo:tool_reward}. This reward encourages the model to explore reasoning that utilizes more of the tools available.

\begin{equation}\label{eq:grpo:tool_reward}
r_{\text{tools}}(x)
=\lambda_{\text{tools}}\cdot \frac{|\mathrm{UniqueToolsUsed}(x)|}{|\mathcal{T}_{\text{avail}}|},
\quad \lambda_{\text{tools}}=0.5,
\end{equation}

This gives a small reward for using a broader set of the provided tools at least once per sample. This nudges the model to query compiler diagnostics and microbenchmarks during reasoning instead of guessing.

\paragraph{Training loop}

\Cref{alg:grpo} details our high-level GRPO training algorithm. For each prompt, we sample a group of $K$ rollouts from $\pi_\theta$, allowing tool calls as part of the reasoning. After each running inference to generate the rollouts we evaluate the performance and correctness of the final generated code. Using these values we compute rewards, normalize them within the group to produce advantages $w(\cdot)$, and apply the GRPO loss with KL regularization with respect to the reference model. The reference is a frozen copy of the original model before any fine-tuning steps. The reward and KL divergence penalty are used to compute the final GRPO loss and, in turn, update the weights of the model using backpropagation and an optimizer step.

\begin{algorithm}[ht]
\caption{GRPO fine-tuning for kernel optimization and tool use}
\label{alg:grpo}
\begin{algorithmic}[1]
\STATE \textbf{Inputs:} dataset $\mathcal{D}$ of optimization tasks \\ baseline times for each task \\ toolset $\mathcal{T}_{\text{avail}}$ \\ main fine-tuning model $\pi_\theta$ \\ reference model $\pi_{\text{ref}}$

\FOR{batch of prompts $c \in \mathcal{D}$}
\STATE For each $c$, sample $K$ rollouts $\{x_k\}_{k=1}^K \sim \pi_\theta(\cdot|c)$ with tools enabled.
\STATE Performance and correctness of final solution in each $x_k$ is evaluated using the benchmark tool.
\STATE Compute $r_{\text{perf}}(x_k)$ and $r_{\text{tools}}(x_k)$; set $r(x_k)=r_{\text{perf}}+r_{\text{tools}}$.
\STATE Normalize $\{r(x_k)\}_{k=1}^K$ within each group to get advantages $w(x_k)$.
\STATE Compute $\mathrm{KL}(\pi_\theta(\cdot|c),|,\pi_{\text{ref}}(\cdot|c))$.
\STATE Take a gradient step on $\mathcal{L}_{\text{GRPO}}$ and update model weights $\theta$.
\ENDFOR
\end{algorithmic}
\end{algorithm}

The main hyperparameters that impact training efficacy in this setup are the total number of training steps taken and the rollout size. Other hyperparameters like learning rate and regularization penalty we observed to have little impact on reward progress during training. We found increasing $K$ for larger rollouts to be the most impactful during fine-tuning, however, at increased memory and compute costs. For this reason we settled on $K=12$ as a good middle ground for large enough rollouts to get good advantage estimates, but also not too computationally intensive.

\subsection{Reasoning Distillation}

Directly running GRPO on small models can be expensive and unstable. Recent work has found that fine-tuning a large model with GRPO and then using the outputs of that model to fine-tune a small model with supervised fine-tuning (SFT) yields much better results than directly running GRPO on the small model~\cite{deepseek-ai_deepseek-r1_2025}. Thus, we first train a larger reasoning model and then distill its traces into a smaller model. Concretely, we (1) run the reasoning model over our dataset, (2) filter for samples that compile, pass correctness, and achieve a speedup greater than one, and (3) perform SFT on the resulting (prompt, reasoning trace, final\_code) tuples. We keep tool-call tokens and arguments in the targets so the small model learns how to call tools. This distillation process produces a small model that retains most of the optimization behavior at a fraction of the inference cost, and performs better than applying GRPO directly to the small model.

\subsection{Fine-tuning Setup}

We implement fine-tuning with HuggingFace Transformers~\cite{wolf_transformers_2020} and PyTorch~\cite{ansel_pytorch_2024}. To distribute training across multiple GPUs and nodes we utilize AxoNN~\cite{singh_axonn_2022} for parallelism, and we run the Yalis~\cite{noauthor_axonn-aiyalis_2025} inference framework on top of AxoNN to compute the GRPO rollouts. We run a single data-parallel training instance and use tensor parallelism to shard both the main policy and the frozen reference model across GPUs. To address tool-latency bottlenecks, we run the training job on a hierarchical set of resources: training steps run on one pool of GPUs while tool calls (builds, microbenchmarks, compiler analyses) are dispatched to a second pool using the Flux scheduler~\cite{ahn_flux_2014,noauthor_230917420_nodate}. Flux queues tool jobs asynchronously from the main training job, so the learner can continue processing rollout inference steps while the tools return their outputs. This engineering design aids in bringing down the training time as we found tool calling to nearly double rollout time, which is already the main bottleneck in GRPO training. By dispatching and asynchronously executing tools we can continue to generate tokens for other members of the rollout group that are not awaiting a tool call output.

As base models for fine-tuning we utilize the Llama3-70b and Llama3-8b models~\cite{grattafiori_llama_2024}.
The larger 70b model is used in the RL training, while its traces are distilled into the smaller 8b model.
We refer to the final fine-tuned models opt-r1 and opt-r1-mini, respectively.

\section{Experimental Setup}\label{sec:setup}
In this section detail our experimental setup: how we evaluate our trained models and the metrics used for comparison.

\subsection{RAJAPerf Benchmarks}

To evaluate and compare the optimization capabilities of the models we utilize the RAJAPerf benchmark suite~\cite{pearce_raja_2025}.
Originally designed to benchmark the performance capabilities of the RAJA library~\cite{noauthor_llnlraja_2025}, the RAJAPerf benchmark suite defines many GPU kernels that it uses as performance comparisons with RAJA kernels.
Many of these kernels are inspired by popular algorithms or real scientific applications and make ideal benchmark candidates.
There are over 70 unique kernels in the RAJAPerf benchmark suite.
Some of these are, however, very trivial (e.g. a memset kernel), so we narrow our focus to the 36 kernels in the application, algorithm, and polybench categories of the benchmark.
The application kernels are GPU kernels inspired directly by real HPC scientific applications.
The algorithm kernels come from common computational algorithms such as reductions and the polybench kernels are taken directly from the PolyBench polyhedral compiler optimization benchmark~\cite{reisinger_matthiasjreisingerpolybenchc-421_2025}.

The RAJAPerf benchmark provides drivers to test both correctness and performance.
Conveniently, RAJAPerf has already handled selecting reasonable problem sizes, number of sample iterations, and launch parameters.
We directly utilize this infrastructure in the benchmark and only have the LM optimize the kernels in the source code.
The only adjustments made were expanding preprocessor macros, since many of the CUDA kernels in the benchmark have their body defined as macros in separate files.
After running the drivers runtimes are directly reported in output result files.

We evaluate both the opt-r1 and opt-r1-mini models on all the kernels in the RAJAPerf benchmark.
Each kernel is optimized five times by the models to account for non-determinism.
They are provided an entire GPU kernel to optimize; in some instances, if that particular benchmark problem spans multiple kernels, we provided multiple kernels to the model to optimize.
As a baseline point of comparison we use the Llama3-70b model~\cite{grattafiori_llama_2024}.
We evaluate it by providing the kernel directly in its prompt and asking it to generate an optimized version of the kernel.

\subsection{Metrics}

To compare approaches and understand how our model is performing we use the speedup metric.
This is the ratio of the original kernel runtime, $t_{\mathrm{orig}}$, to the runtime of the proposed optimized kernel, $t_{\mathrm{opt}}$, as shown in \Cref{eq:speedup}.

\begin{equation}\label{eq:speedup}
    \mathrm{speedup}=\frac{t_{\mathrm{orig}}}{t_{\mathrm{opt}}}
\end{equation}

This provides information on how well the model is able to decrease the runtime of the GPU kernel.

\section{Results}\label{sec:results}
In this section we detail the evaluation of the opt-r1 and opt-r1-mini models on the RAJAPerf GPU kernels. We further examine reasoning traces and ablation studies to better understand how the reasoning model approaches optimization.

\subsection{Optimizing RAJAPerf Kernels}

\begin{figure}[ht]
    \centering
    \includegraphics[width=0.98\linewidth]{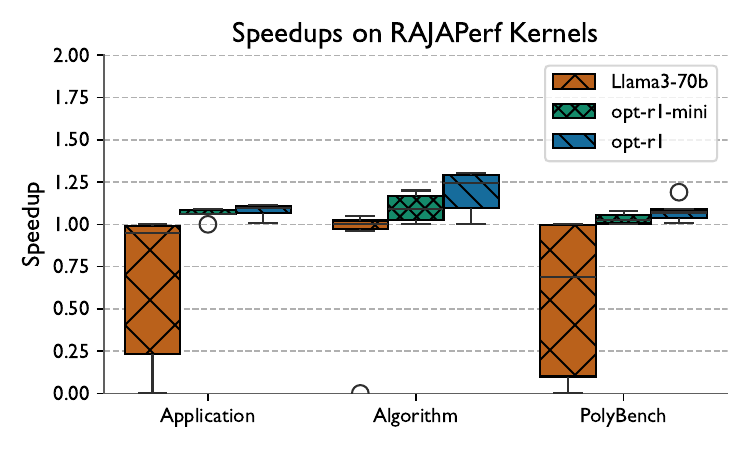}
    \caption{The distribution of speedups for each model across the kernel types in the RAJAPerf suite. A speedup of 0 denotes a failed result (i.e. incorrect or has syntax errors). The opt-r1 and opt-r1-mini models are consistently able to improve kernel performance across most benchmark problems. \label{fig:speedups}}
\end{figure}

We evaluate the models on the task of optimizing CUDA kernels from the RAJAPerf benchmark suite.
\Cref{fig:speedups} shows the range of speedup values obtained by each model on the kernels in each evaluated category of RAJAPerf. 
Llama3-70b is included as a baseline.
Speedup numbers are obtained using the RAJAPerf drivers to get the original and modified kernel performance.
We additionally test to ensure the generated kernel is correct. If it is incorrect or does not compile, then we denote the speedup as 0.

We see several important trends in the speedup results. First, the optimization reasoning models consistently achieve speedups greater than one, whereas the vanilla Llama3-70b model sometimes produces kernels that are incorrect or slower than the original kernel. This is due to the reasoning models being able to test hypotheses and iterate on code samples; if an optimization attempt does not work, then it will move on to the next one. Sometimes the opt-r1 models achieve a speedup of 1.0 meaning they could not find an optimization that works, so they return the original kernel. This is incredibly useful in comparison to the Llama3-70b model, which sometimes yields invalid outputs.

A second, important trend is that the opt-r1 model yields better results than the opt-r1-mini model on average. The opt-r1-mini is still able to generate successful kernel optimizations, which is impressive due to its size. This model is easier to run and could be practically deployed on a smaller GPU or CPU.

\subsection{Examining Progress Throughout Traces}

\begin{figure}[ht]
    \centering
    \includegraphics[width=0.98\linewidth]{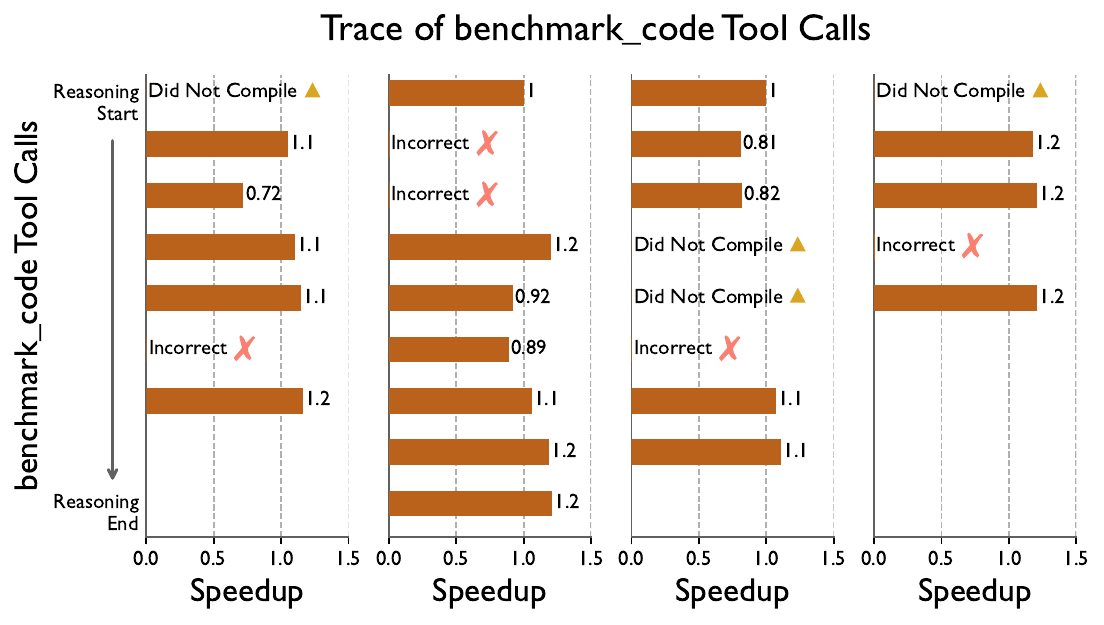}
    \caption{Four traces of \texttt{benchmark\_code} tool call results during optimization of the FEMSweep RAJAPerf kernel by opt-r1.
    Each plot shows the speedup of progressive invocations of the benchmark tool during the model's reasoning process.
    \label{fig:tool-trace}}
\end{figure}

\Cref{fig:tool-trace} displays speedups from invocations of the \verb|benchmark_code| tool during reasoning across four different runs of opt-r1 on the FEMSWEEP kernel.
Labels are used to denote invocations that did not compile or were incorrect.
We observe several interesting trends: (1) performance does not gradually increase as the model tries new optimizations, (2) the model's strategy varies from run to run, and (3) the model sometimes produces invalid optimizations (i.e. does not compile or is incorrect).

The first insight, which gives insight into how the model approaches problems, is that there is no strong trend in the performance of code the model attempts to benchmark. One might anticipate that the model gradually tries more optimizations and the code gets faster and faster over time. However, a common trend we see in the data is that the model finds a valid optimization early on and then proceeds to attempt more aggressive optimizations. If these do not work out because they are slower or incorrect, then the model will come back to its earlier attempt. This trend is displayed in \Cref{fig:tool-trace} where the model frequently achieves a good speedup, then produces invalid or poor performing code, and finally finishes with an optimized kernel as fast as its earlier attempts.

\Cref{fig:tool-trace} further demonstrates the indeterminacy of the model. While each trace is different due to randomness from output sampling and tool outputs, we see that in all cases the model is able to provide an optimized output. The outputs are successfully, but we cannot expect consistency in the types of solutions the model derives. Among these varying solutions we also observe that the model sometimes produces code that is incorrect or invalid. This highlights the necessity of the tool calling reasoning model that can dynamically interact with the environment and retry optimizations after failed attempts. If the model is not allowed to continue reasoning about optimization strategies, then it will yield invalid final solutions a non-negligible amount of attempts.

\subsection{Tool Ablation Study}

\begin{figure}[ht]
    \centering
    \includegraphics[width=0.98\linewidth]{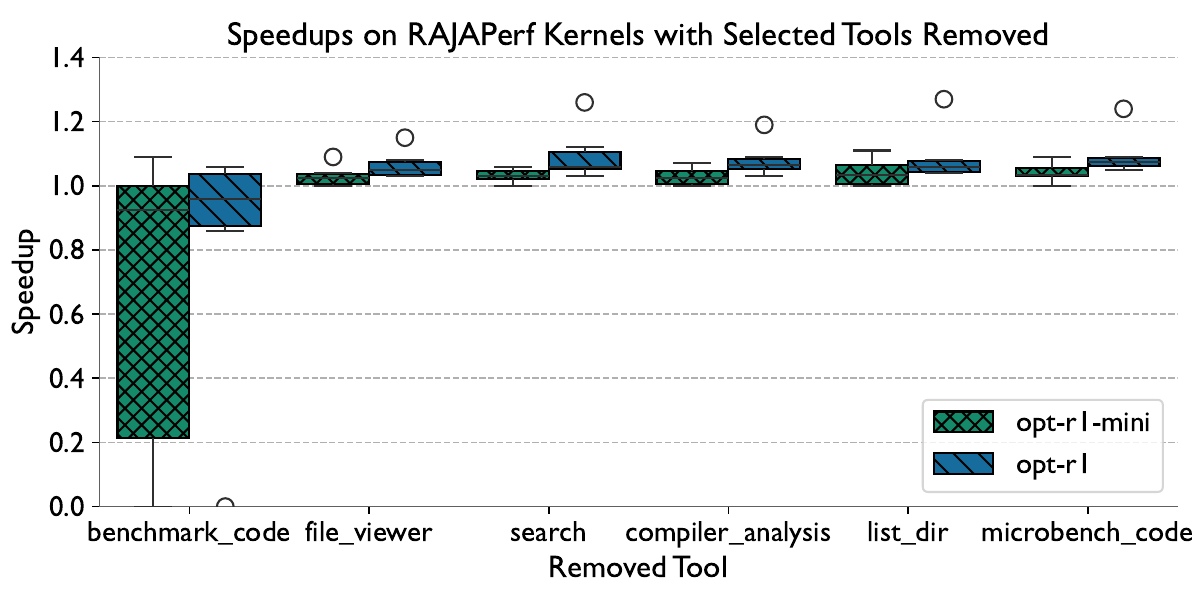}
    \caption{The distribution of speedups on the RAJAPerf kernels from opt-r1 and opt-r1-mini with specified tools removed. For example, the last bar shows the distribution of speedups where the \texttt{microbench\_code} tool is not available. We observe that the \texttt{benchmark\_code} tool is crucial to consistently getting valid optimizations from the models. \label{fig:tool-ablation}}
\end{figure}

To better understand the impact of each tool available we conduct an ablation study by optimizing the RAJAPerf kernels with each tool removed. \Cref{fig:tool-ablation} shows the speedup numbers on the benchmark with each tool removed (i.e. the model has five tools available with one omitted).

We immediately see that the \texttt{benchmark\_code} tool is critical to obtaining kernel speedups from the model.
Without it the model is not able to test its optimizations or see if the code compiles.
This leads to the opt-r1 models sometimes produces incorrect and/or non-compilable code in their final outputs.
We see little variation with the other tools.
While they all provide added context and capabilities to the model, none of them, if removed, leads to a drastic collapse in performance.

\begin{figure}[ht]
    \centering
    \includegraphics[width=0.98\linewidth]{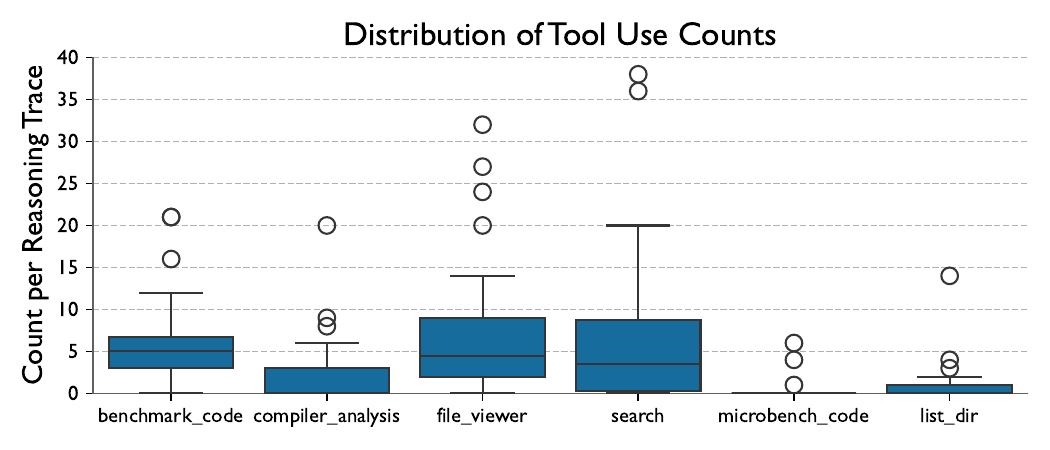}
    \caption{Tool use counts per reasoning trace across all traces from the RAJAPerf evaluation. \texttt{benchmark\_code}, \texttt{file\_viewer}, and \texttt{search} are the most widely used, while most traces rarely use the \texttt{microbench\_code} tool more than once. \label{fig:tool-use}}
\end{figure}

Due to the models not being fine-tuned with data samples where only one tool was available, we do not perform this experiment with only one of the six tools available to the model.
Instead we qualitatively analyze reasoning traces and observe many instances of the model using the other tools to find valuable insights during their optimization process.
Furthermore, \Cref{fig:tool-use} shows the count of tools used in each trace.
We see the model consistently using the tools during its reasoning with the exception of the \texttt{microbench\_code} tool, which is used sparingly.

\section{Case Study: Optimizing a Shock-Hydrodynamics Kernel on H100}\label{sec:case-study}
In this section we present a case study using our optimization model to optimize an expensive kernel in a shock hydrodynamics code proxy application, Lulesh~\cite{noauthor_llnllulesh_2025}, and achieve a 3\% improvement in total application performance.

\subsection{Kernel Description}
We target the ``CalcKinematicsAndMonotonicQGradient\_kernel", which computes updated geometric and kinematic quantities for each mesh element in a 3D Lagrangian hydrodynamics simulation. The kernel processes one element per thread, updating properties such as element volume, deformation, strain rates, and monotonic gradients used for artificial viscosity.
After profiling Lulesh we found this to be the second most expensive kernel in the code and typically very close in runtime to the slowest kernel making it a good optimization target.

\subsection{Reasoning Trace}

The reasoning trace produced by the model for this kernel is significant in length at several thousand lines of text. For the sake of brevity we summarize the reasoning process in \Cref{tab:case-study:reasoning-trace}. The model first hypothesizes about three optimization strategies; it tries them, but the benchmark tool fails to compile. After this the model makes several changes and calls to the tools to figure out how to get the code to compile. Once the code is compiling and correct it then tries several different compiler flags to see how they impact performance and correctness.
There are several other strategies the model tries, but throws out after benchmarking and determining they are not helpful.
Finally, it returns a valid version of the kernel with optimizations that lead to a 3\% improvement in total application time. 

\begin{table*}[th]
\centering
\caption{Model’s iterative reasoning loop: hypothesize \(\rightarrow\) edit \(\rightarrow\) compile \(\rightarrow\) measure \(\rightarrow\) iterate.}
\label{tab:case-study:reasoning-trace}
\begin{tabular}{@{}p{4.1cm}p{5.6cm}p{2.2cm}@{}}
\toprule
\textbf{Hypothesis / Plan} & \textbf{Action \& Observation} & \textbf{Outcome} \\
\midrule
Establish baseline & Build default (\texttt{-O3, sm\_90}); run benchmark. & 4.29–4.30 s \\
\addlinespace[2pt]
ROI is load-heavy; reduce memory traffic and loop overhead. & Patch kernel to (i) use \verb|__ldg| for read-only arrays \{\texttt{x,y,z,xd,yd,zd,volo,v}\}; (ii) fuse node/coord loads in one unrolled loop; (iii) avoid “undo” pass by computing adjusted coords and preserving originals. First compile \emph{fails}: missing host inline wrapper at call site. & \ding{55} build error \\
\addlinespace[2pt]
Fix integration issues. & Retry patch including the unmodified host inline wrapper that launches the kernel; verify symbol name/call site. Recompile. & \checkmark{} builds \\
\addlinespace[2pt]
Try modest compiler flag changes. & Build with \texttt{-use\_fast\_math} and target \texttt{sm\_90a} (last \texttt{-arch} wins); correctness checks pass. & \checkmark{} correct \\
\addlinespace[2pt]
Measure effect. & Run end-to-end benchmark on $120^3$. & 4.17 s \\
\bottomrule
\end{tabular}
\end{table*}

\vspace{0.6em}
\noindent\emph{Considered \& dropped.} The model analyzed changing \verb|__launch_bounds__| and larger blocks, using shared memory for per-thread temporary values, and prefetch/vectorized loads; it abandoned these strategies after noting the code's read-only, gather-heavy pattern and acceptable register usage (no spills), focusing instead on reducing passes and leveraging the read-only cache.

\subsection{Model Optimizations}
The final optimizations yielded by the model were:

\begin{enumerate}[leftmargin=*,itemsep=2pt]
  \item \textbf{Read-only caching:} Use \verb|__ldg| for \texttt{x,y,z,xd,yd,zd,volo,v}.
  \item \textbf{Loop fusion:} Fuse node-index and coordinate loads into one unrolled loop to reduce overhead and improve ILP.
  \item \textbf{Single-pass coordinates:} Compute adjusted half-step coordinates for kinematics while preserving original coordinates for the monotonic \(Q\) gradient, eliminating the extra “undo” loop.
  \item \textbf{Build flags:} Compile with \texttt{-use\_fast\_math} and target \texttt{sm\_90a}.
\end{enumerate}

Each of these four strategies were employed in the kernel and led to reduced run time. We found that the \verb|__ldg| intrinsic was the most common optimization strategy the model used across all of our benchmarking; this is a low-hanging optimization that most GPU codes omit.
After attempting the \verb|__ldg| intrinsic optimization the model noticed that two loops could be fused and part of the coordinate projection logic could be modified to prevent having to ``undo" computation at the end of the kernel.
The model found these optimizations, combined with ideal build flags, to improve the end-to-end time.

\subsection{Results}

The time spent in the ``CalcKinematicsAndMonotonicQGradient\_kernel" for problem size $120^3$ is reduced by 17\%. This yields a 3\% improvement in performance for the entire application. These results are displayed in \Cref{tab:case-study:lulesh-results} for executions on an NVIDIA H100 GPU.

\begin{table*}[th]
\centering
\caption{Kernel and end-to-end wall time on NVIDIA H100 (problem size: $120^3$). Correctness checks passed in both runs.}
\label{tab:case-study:lulesh-results}
\begin{tabular}{lrrr}
    \toprule
    Configuration & Kernel Time (s) & Time (s) \\
    \midrule
    Baseline (\texttt{-O3, sm\_90}) & 0.881 & 4.29–4.30 \\
    Optimized (edits + \texttt{-use\_fast\_math, sm\_90a}) & 0.75 ({\raise.17ex\hbox{$\scriptstyle\sim$}}17\%) & 4.16-4.17 ({\raise.17ex\hbox{$\scriptstyle\sim$}}3\%) \\
    \bottomrule
\end{tabular}
\end{table*}

\section{Related Work}\label{sec:related-work}
In this section we detail related works on using LMs for code and GPU kernel optimization.

Several recent works have begun looking into utilizing LMs to automate code optimization.
With the rise of LMs and their intense compute requirements, there has been a particular interest in optimizing GPU kernels.
One thread of recent work, most similar to this work, looks into using RL to train optimization models.
RL is amenable to optimization as the task has a natural {\it verifiable reward signal} to use during fine-tuning.
Nichols et al. introduce an RL methodology to align code LLM outputs with execution speed, achieving generated code speedups of up to $1.6\times$ on serial benchmarks and $4.5\times$ on parallel OpenMP code~\cite{nichols_performance-aligned_2024}.
While similar to ours, this work merely aligns the output distribution of a LM with fast code making it more likely to generate fast code. However, the model does not have dynamic feedback through a reasoning process and tool use.
Other similar works employ RL for optimization~\cite{duan_perfrl_2025,tschand_swizzleperf_2025,andrews_gpu_2025}, but none train reasoning models and/or models that can dynamically interact with performance tools.
We refer the reader to \cite{gong_language_2025} for a survey of works in this area.

Several works have focused more on collecting high quality optimization text data and simply using supervised fine-tuning, rather than RL~\cite{shypula_learning_2024,garg_deepdev-perf_2022,shypula_learning_2024,huang_efficoder_2025}. 
Garg et al.~\cite{garg_deepdev-perf_2022} use mined git commits related to performance to fine-tune a model to generate performance improving C\# code edits.
Huang et al.~\cite{huang_efficoder_2025} develop a synthetic data generation pipeline to create high quality optimization samples and fine-tune LMs on this dataset. Further works have explored benchmarking LM generated code performance as a stepping stone to optimization~\cite{nichols_can_2024,ouyang_kernelbench_2025}, but many of these focus on specific types of tasks, e.g. mapping PyTorch code to CUDA kernels~\cite{ouyang_kernelbench_2025}.

\section{Discussion and Limitations}\label{sec:discussion}
This section discusses the constraints of our work and how it can fit in the broader picture of LM guided code optimization.

\subsection{Optimization Scope and Broader Optimization Agents}

By design, our work narrows itself in focus to the well-scoped problem of optimizing a contiguous region of GPU source code in a repository when correctness checks and performance benchmarks are readily available. The constraints of this choice, while limitations, also deeply strengthen the capabilities of the model. Constraining edits to a fixed, small portion of the code and requiring verifiable tests focus the model on a concrete task and turn ``performance" into an easily measurable signal. The model is faced with a much more focused, feasible task. The trade-off is that some impactful end-to-end optimizations may lie outside the code boundary (e.g., refactoring data layouts or changing a global variable). In practice, however, smaller scoped optimization problems are useful primitives that can be composed into broader workflows, likely as LM-powered optimization agents. The models proposed in this work can be easily integrated as subcomponents in a larger optimization framework; they focus on optimizing small sets of kernels while the larger code orchestrates repository-level optimizations. Large scale agentic approaches to optimization are out of the scope of this work, but serve as promising future directions.

There are two other major limiting factors in our approach: the context window and the latency budget for test-time compute and tool calls. Long target kernels and reasoning traces can push token limits; the model has a fixed length context window and can only process so many tokens at once. Given that the model will receive poor rewards during training if it is incapable of finding an optimization in its context window length, this outcome should be unlikely in the final model. Furthermore, we observed in our results and case study that the model was able to find successful optimizations in its context window. Thus, while the context length remains a limitation, we consider it minor in practice. The other major limitation is the increased optimization time. Running inference with the LM for long reasoning traces and expensive tool calls can take 5 to 20 minutes and requires available GPU hardware to serve the model and run benchmarks. While this does impose more time and hardware constraints than traditional compiler optimizations we consider it likely that developers will accept a one-time 5 to 20 minute optimization run to increase their overall performance by a few percentage points.

\subsection{Language Model Guided Code Optimization}

LMs for code optimization, particularly GPU kernel optimization, has the potential for great impact on software development and high performance computing. 
Performance engineering is a traditionally difficult task often left to experts.
Putting the capability to rapidly explore and discover source-level code optimizations into the hands of domain scientists and code developers will drastically increase the rate at which fast code and results can be delivered.
This work provides a crucial stepping stone towards fully automated code optimization.
It is important that these future works consider the { proper computer interface for environment interaction}, { how performance engineering shifts into constraint engineering}, and { expensive tool calling}.

\noindent\textit{Computer Interface for Optimization.} 
In the same manner that human-centered software requires good user interfaces, an important design consideration for LM-based optimization frameworks is the ``computer interface". Adding more tools is not automatically helpful; in our experimentation we observed that models struggle with complex, loosely specified APIs and benefit from a small number of reliable tools aligned to common performance queries (compile status, correctness, timing, register/occupancy hints, simple search). 

\noindent\textit{Performance Engineering as Constraint Engineering}
Another important implication from this work is the transition of performance engineering problems into constraint engineering.
LMs are becoming more proficient at handling complex software engineering tasks, but still struggle at grasping the larger context of problems they are solving. 
For example, a LM will not know what codes need to be optimized, where and how it can run them, what problem sizes are of relevant interests to scientists, etc.
As the LMs get better at solving complex problems, like kernel optimization, it shifts the human labor to defining problem constraints for the LM.

\noindent\textit{Expensive Tool Calling.}
Finally, an important consideration for LM optimization frameworks is the readiness of the software stack for these types of workloads. One particular pain point found in this work is the lack of stable solutions for expensive tool calling. Many inference and agentic software are designed with tools that execute in a few seconds in mind. They do not make any optimizations for tools that take several minutes or be a complex workflow of stages.

\section{Conclusion}\label{sec:conclusion}
This work explores the hypothesis that allowing reasoning LMs to 
interact with execution environments during their reasoning can greatly enhance their capabilities on tasks like code optimization. Specifically, we explore how a large LM can be fine-tuned to reason about GPU kernel optimization and utilize performance optimization tools during its reasoning. This is accomplished by exposing a compact toolset to the language model (compiler diagnostics, correctness checks, and benchmarks) that it can utilize as it reasons about how to optimize a GPU kernel.

We then demonstrated how GRPO and RL can be utilized to fine-tune a reasoning optimization model to use tools during its reasoning. It was further demonstrated that this reasoning capability can be distilled into a smaller LM using supervised fine-tuning. These models are demonstrated to improve kernel performance on realistic application inspired GPU kernels in the HeCBench benchmark. It is further shown that the introduced models can optimize kernels from real applications; we are able to improve end-to-end performance of a shock-hydrodynamics application by 3\%.
Importantly, these optimizations are not just code edits, but LM generations that are accompanied by natural language explanations and synthesis.

\section*{Acknowledgment}
This work was performed under the auspices
of the U.S.~Department of Energy by Lawrence Livermore National Laboratory
(LLNL) under Contract DE-AC52-07NA27344 (LLNL-JRNL-2012719). This work was
supported in part by LLNL LDRD projects 25-ERD-058.
This material is based upon work supported by the U.S. Department of Energy, Office of Science, Office of Advanced Scientific Computing Research, through solicitation DE-FOA-0003264, “Advancements in Artificial Intelligence for Science,” under Award Number DE-SC0025598 and contract DE-AC52-07NA27344.

\bibliographystyle{ieeetr}
\bibliography{bib/external-ref,bib/zotero-references}

\end{document}